\documentclass[allclo]{FBSart}
\usepackage{amsfonts}
\usepackage{amsmath}
\usepackage{amssymb}
\usepackage{multirow}
\usepackage{bm,graphicx}
\usepackage{wrapfig}

\def\a{\alpha}

\title{Selected Topics in Three- and Four-Nucleon Systems
\footnote{Presented at the 21st European Conference on Few-Body
Problems in Physics, Salamanca, Spain, 30 August - 3 September 2010.}}
 \author{A. Kievsky\instnr{1}\thanks{alejandro.kievsky@pi.infn.it} }
\instlist{ Istituto Nazionale di Fisica Nucleare,
          Largo Pontecorvo 3, 56100 Pisa, Italy.}
\runningauthor{A. Kievsky}
\runningtitle{Selected Topics in Three- and Four-Nucleon Systems}

\begin{document}
\maketitle
\abstract{
Two different aspects of the description of three- and four-nucleon systems
are addressed. The use of bound state like wave functions to
describe scattering states in $N-d$ collisions at low energies and the
effects of some of the widely used three-nucleon force models in 
selected polarization observables in the three- and four-nucleon systems
are discussed.
} 
\maketitle
%
%
%
\section{Introduction}
\label{kievsky_intro}

Detailed studies in the three- and four-nucleon systems gives valuable
information of the underlying nuclear interaction. These two systems
have three bound states, $^3$H, $^3$He and $^4$He, therefore much of the
efforts have been done in the study of continuum states. Although
a reasonable agreement with the available experimental data
is obtained in the description of the differential
cross section in the low energy region, discrepancies can be observed in
some polarization observables~\cite{kiev01,fisher06}. Related to this,
the analysis of the effects of the three-nucleon forces are of
crucial importance. Recently a critical comparison of different
models widely used in the literature has been performed~\cite{kiev10b}.

A different aspect of the problem regards the methods used to describe
continuum states in few-nucleon systems. In the $A=3,4$ systems well
established methods to treat both, bound and scattering states,
are the solution of the Faddeev equations ($A=3$) or Faddeev-Yakubovsky 
equations ($A=4$) in configuration or momentum space and the Hyperspherical 
Harmonic (HH)
expansion in conjunction with the Kohn Variational Principle (KVP). These
methods have proven to be of great accuracy and they have been tested through
different benchmarks~\cite{benchmark1,benchmark2}.
On the other hand, other methods
are presently used to describe bound states: for example the Green
Function Montecarlo (GFMC) and No Core Shell Model (NCSM) methods
have been used in nuclei up to $A=10$ and $A=12$ respectively~\cite{gfmc,ncsm}.
The possibility of employing bound state techniques to describe scattering states
has always attracted particular attention. Recently
continuum-discretized states obtained from the stochastic variational method
have been used to study $\alpha+n$ scattering~\cite{suzuki}. In a different
approach continuum states have been obtained using bound state like
wave functions~\cite{kiev10a}.

In the present paper we discuss the description of $N-d$ scattering states 
using bound state like wave functions and we 
briefly show three-body force effects in selected polarization observables
in $A=3,4$.

\section{Continuum states from bound state like wave functions}
\label{sec:sec1}

 Following Refs.~\cite{kiev10a,barletta09}, it was shown that a second order
estimate of the scattering matrix $R$ at a collision energy $E$ 
(below the breakup threshold) results
\begin{equation}
\left.
\begin{array}{ccc}
B^{2^{nd}}_{ij}&=&-<\Psi_i|H-E |{\cal F}_j> \\
A_{ij}&=& <\Psi_i|H-E|{\cal G}_j>
\end{array} \right\}
R^{2^{nd}}=A^{-1}B^{2^{nd}}.
\label{eq:second}
\end{equation}
The eigenvalues of $R^{2^{nd}}$ are second order estimates of the
phase shifts and the indeces $(i,j)$ indicate
the different asymptotic configurations accessible at the specific energy
under consideration. In particular, for the three-nucleon system,
${\cal F}_j$ and ${\cal G}_j$ are the channel wave functions describing the
possible relative states of the deuteron and the incident nucleon. 
For a given $J^\pi$ state
the different channels are labelled by the relative angular momentum $L$ between
the deuteron and the incoming nucleon coupled to the total spin $S=1/2$ or $3/2$
obtained coupling the spin $s_d=1$ of the deuteron the the sin $s=1/2$ of the
incoming nucleon. Specifically $j\equiv L,S,J$ and the channel functions are

\begin{eqnarray}
   {\cal F}_{LSJ} = \sum_i\left[ \sum_{l_\a=0,2} w_{l_\a}(x_i)
       F_L (y_i)
       \left\{\left[ [Y_{l_\a}({\hat x}_i) s_\a^{jk}]_1 s^i \right]_S
        Y_L({\hat y}_i) \right\}_{JJ_z}
       [t_\a^{jk}t^i]_{TT_z}\right] \ ,  \\
   {\cal G}_{LSJ} = \sum_i\left[ \sum_{l_\a=0,2} w_{l_\a}(x_i)
       \tilde G_L (y_i)
       \left\{\left[ [Y_{l_\a}({\hat x}_i) s_\a^{jk}]_1 s^i \right]_S
        Y_L({\hat y}_i) \right\}_{JJ_z}[t_\a^{jk}t^i]_{TT_z}\right] 
\label{eq:asymp}
\end{eqnarray}
The sum on $i$ runs over the three cyclic permutations of the jacobi coordinates,
$x_i,y_i$ are their moduli, ${\hat x}_i,{\hat y}_i$
their directions and
$w_l(x_i)$ the $l$-wave deuteron wave function. The functions $F_L(y_i)$
and $\tilde G_L(y_i)$ are the regular and irregular solutions of the $N-d$
Schr\"odinger equation outside the interaction region.
The irregular solution has been opportunely regularized at the origin as
\begin{equation}
       \tilde G_L (y)=(1-{\rm e}^{-\gamma r_{Nd}})^{L+1} G_L(y)
\label{eq:reg}
\end{equation}
where $r_{Nd}=(\sqrt{3}/2)\;y$ is the nucleon-deuteron separation and $\gamma$
a parameter that is fixed requiring that $\tilde G_L (y)\equiv G_L(y)$
asymptotically. Moreover, $F_L,G_L$ are the regular and
irregular Bessel functions or Coulomb functions in the case of
$n-d$ or $p-d$ scattering, respectively.

\begin{figure}
\includegraphics[width=7cm,height=6cm,angle=0]{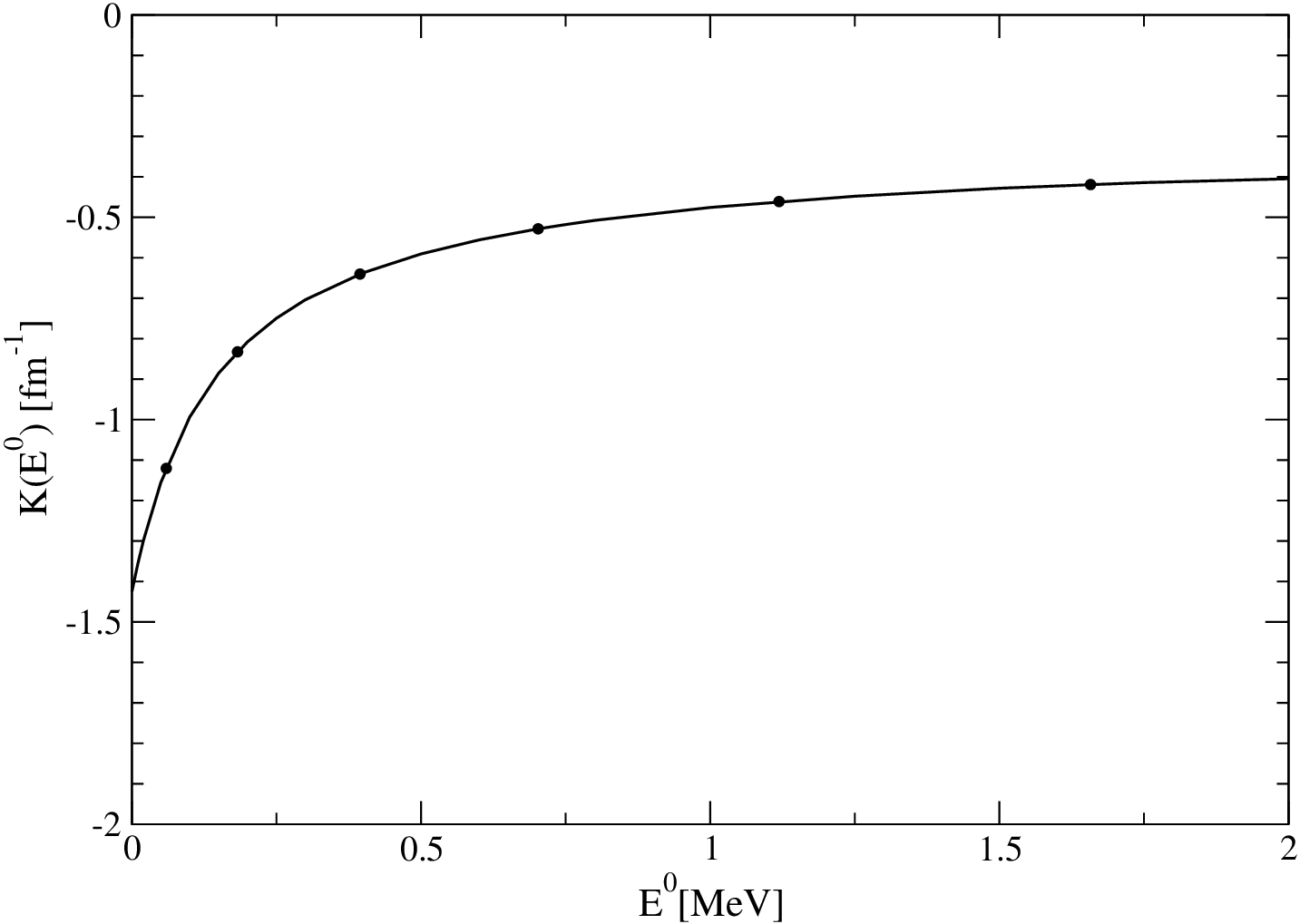}\hspace{0.2cm}
\includegraphics[width=7cm,height=6cm,angle=0]{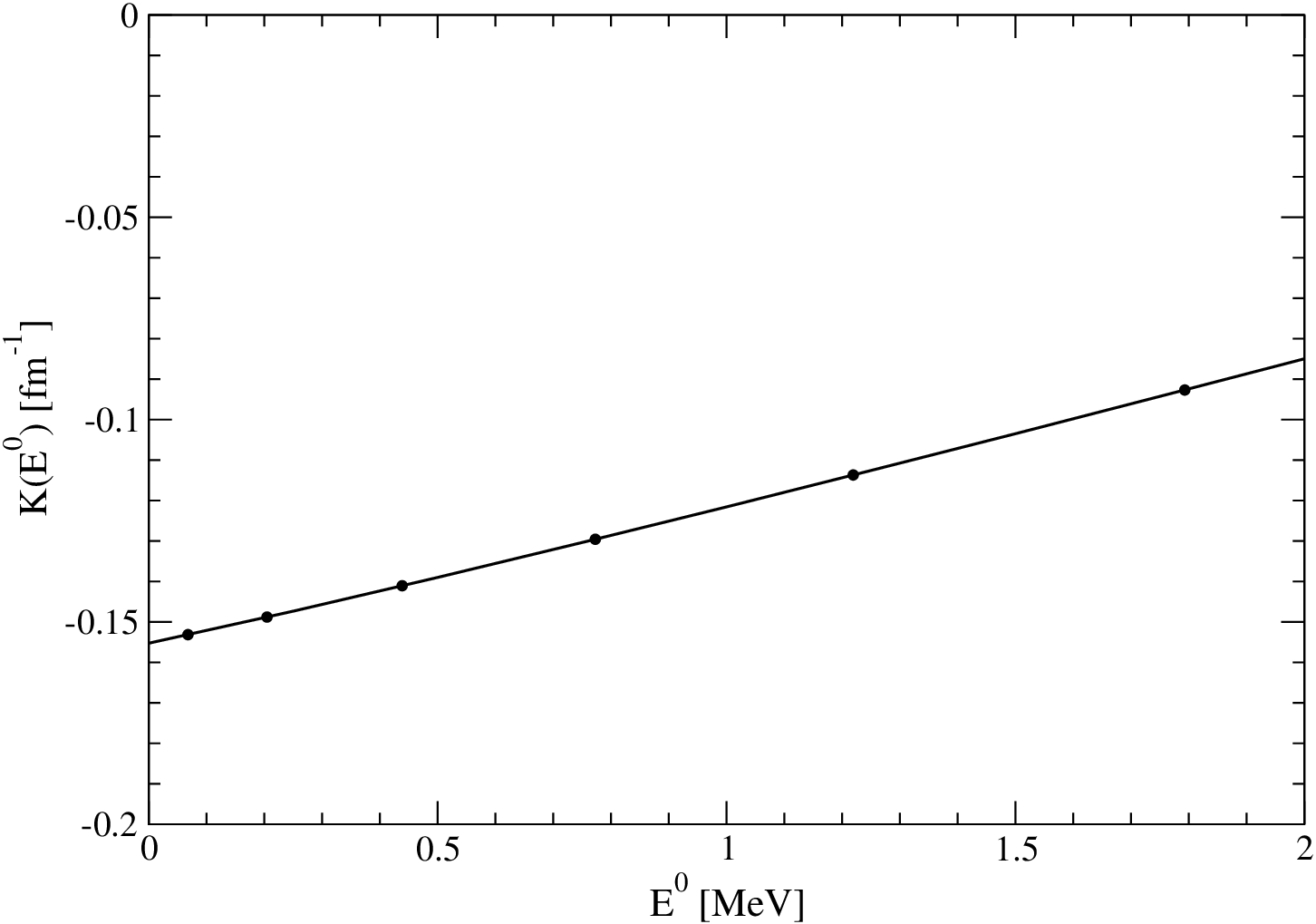}\hspace{0.2cm}
\caption{The $n-d$ effective range function for $J=1/2^+$ (left panel) and $J=3/2^+$
(right panel). The solid points are obtained from the
second order estimates using the integral relations with bound state like
wave functions at the corresponding energies.}
\label{fig:fig1}
\end{figure}

The relations given in Eq.(\ref{eq:second}) are derived from the KVP,
formulating it in terms of integral relations depending 
on the internal structure of the wave function $\Psi_i$. In fact
$(H-E){\cal F}_j$ and $(H-E){\cal G}_j$ go to zero in the asymptotic region
since ${\cal F}_j,{\cal G}_j$ are the solutions of $(H-E){\cal F}_j,{\cal G}_j=0$ 
in that limit. Therefore,
in Eq.(\ref{eq:second}), it would be possible to use trial wave functions $\Psi_i$
that are solutions of $(H-E)\Psi_i=0$ in the interaction region but do not have 
the physical asymptotic behavior indicated in Eq.(\ref{eq:asymp}). 
In particular, it would be
possible to use the bound state like wave functions which are solutions of 
$(H-E_n)\Psi^{(n)}=0$ in the interaction region at particular values of the
energy $E_n$. To explore this possibility,
let us define a complete square integrable basis $|J^\pi,\alpha>$
to expand a bound state like wave function corresponding to a state having total
angular momentum and parity $J^\pi$,
\begin{equation}
       \Psi^{(n)}=\sum_\alpha A^n_\alpha |J^\pi,\alpha>
\label{eq:bound1}
\end{equation}
The index $\alpha$ indicates all the quantum numbers necessary to define the state
and the linear coefficients of the expansion can be obtained from the following
generalized eigenvalue problem
\begin{equation}
    \sum_{\alpha'} A^n_{\alpha'} <J^\pi,\alpha|H-E_n|J^\pi,\alpha'>=0 \;\; .
\label{eq:geneig}
\end{equation}
For example, considering the state $J^\pi=1/2^+$ of the three-nucleon system,
the lowest eigenvalue after the diagonalization procedure 
corresponds to the three-nucleon bound state energy of $^3$H ($T_z=-1/2$) or
$^3$He ($T_z=1/2$). However, as shown in Ref.~\cite{kiev10a},
more negatives eigenvalues could appear
verifying $|E_n|< |E_d|$, with $E_d$ the deuteron binding energy.
The corresponding eigenvectors $\Psi^{(n)}$ approximately describe a scattering process
at the center of mass energy $E_n^0=E_n-E_d$, though asymptotically they
go to zero. Considering other $J^\pi$ states, the diagonalization procedure
will not produce bound states since, in the three-nucleon system, a bound state
exists only in the $J^\pi=1/2^+$ state.
However negative eigenvalues could appear, verifying $|E_n|< |E_d|$. 
As in the previous
case, the corresponding eigenvectors approximately describe $N-d$ scattering states,
though asymptotically they go to zero. 
The eigenvalues $E_n$ are embedded
in the continuum spectrum of $H$ which starts at $E_d$. Accordingly,
increasing the dimension of the basis the number of them increases. We can
consider these states approximate solutions of $(H-E_n)\Psi^{(n)}=0$ in the interaction
region and use them as inputs in the integral relation to compute
second order estimate of the phase-shifts. As an example, results for scattering
$J=1/2^+,3/2^+$ states are given in Fig.~\ref{fig:fig1} using the
$s$-wave MT I-III nucleon-nucleon interaction~\cite{mtiii}. 
The $n-d$, $l=0$, phase shifts $\delta$ are given as a function of the
energy in form of the effective range functions.
For $n-d$ scattering this function is defined as 
$K(E^0)=k\cot\delta$, with $E^0=E-E_d$. 
The solid line in the figures represents this function 
in the interval $[0,|E_d|]$. The solid points
in the figures are the results obtained from the integral relations using
bound state like wave functions at the corresponding energies. As can be observed,
the results using the bound state like wave functions are in complete agreement
with the exact results given by the solid line in all the energy interval.

\section{Analysis of Three Nucleon Force Models}
\label{sec:sec2}

In order to reproduce correctly the three-nucleon bound state energy,
different three-nucleon force (TNF) models have been constructed during the
past years as the Tucson-Melbourne (TM), Brazil (BR) and the
Urbana IX (URIX) models \cite{tm,brazil,urbana}. These models are based
on the exchange mechanism of two pions between three nucleons.
More recently, TNFs have been derived~\cite{epelbaum02}
using a chiral effective field theory at next-to-next-to-leading order.
A local version of these interactions (hereafter referred as N2LOL)
can be found in Ref.~\cite{N2LO}. At next-to-next-to-leading order, the TNF has
two unknown constants that have to be determined. It is a common
practice to determine these parameters from the three- and four-nucleon binding
energies ($B$($^3$H) and $B$($^4$He), respectively).

The $n-d$ doublet scattering length $^2a_{nd}$
is correlated, to some extent, to the $A=3$ binding
energy through the so-called Phillips line~\cite{phillips,bedaque}.
However the presence of TNFs could break this correlation.
Therefore $^2a_{nd}$ can be used as an independent
observable to evaluate the capability of the interaction models to
describe the low energy region.
In Ref.~\cite{report} results for different combinations of NN interactions
plus TNF models are given. These results are shown for the quantities of
interest in Table I and
are compared to the experimental values of the binding
energies and $^2a_{nd}$~\cite{doublet}.
From the table, we can observe that
the models are not able to describe simultaneously the $A=3,4$
binding energies and $^2a_{nd}$. 

\begin{table}[h]
\caption{The triton and $^4$He binding energies $B$ (MeV),
and  doublet scattering length $^2a_{nd}$ (fm)
calculated using the AV18 and the N3LO-Idaho
two-nucleon potentials, and
the AV18+URIX, AV18+TM' and N3LO-Idaho+N2LOL two- and three-nucleon interactions.
The experimental values are given in the last row.}
\label{tb:table1}
\begin{tabular}{@{}llll}
\hline
Potential & $B$($^3$H) & $B$($^4$He) & $^2a_{nd}$ \cr
\hline
AV18            & 7.624    & 24.22   & 1.258 \cr
N3LO-Idaho      & 7.854    & 25.38   & 1.100 \cr
AV18+TM'        & 8.440    & 28.31   & 0.623 \cr
AV18+URIX       & 8.479    & 28.48   & 0.578 \cr
N3LO-Idaho+N2LOL & 8.474    & 28.37   & 0.675 \cr
\hline
Exp.            & 8.482    & 28.30   & 0.645$\pm$0.003$\pm$0.007 \cr
\hline
\end{tabular}
\end{table}

\begin{figure}
\includegraphics[width=14cm,height=6cm,angle=0]{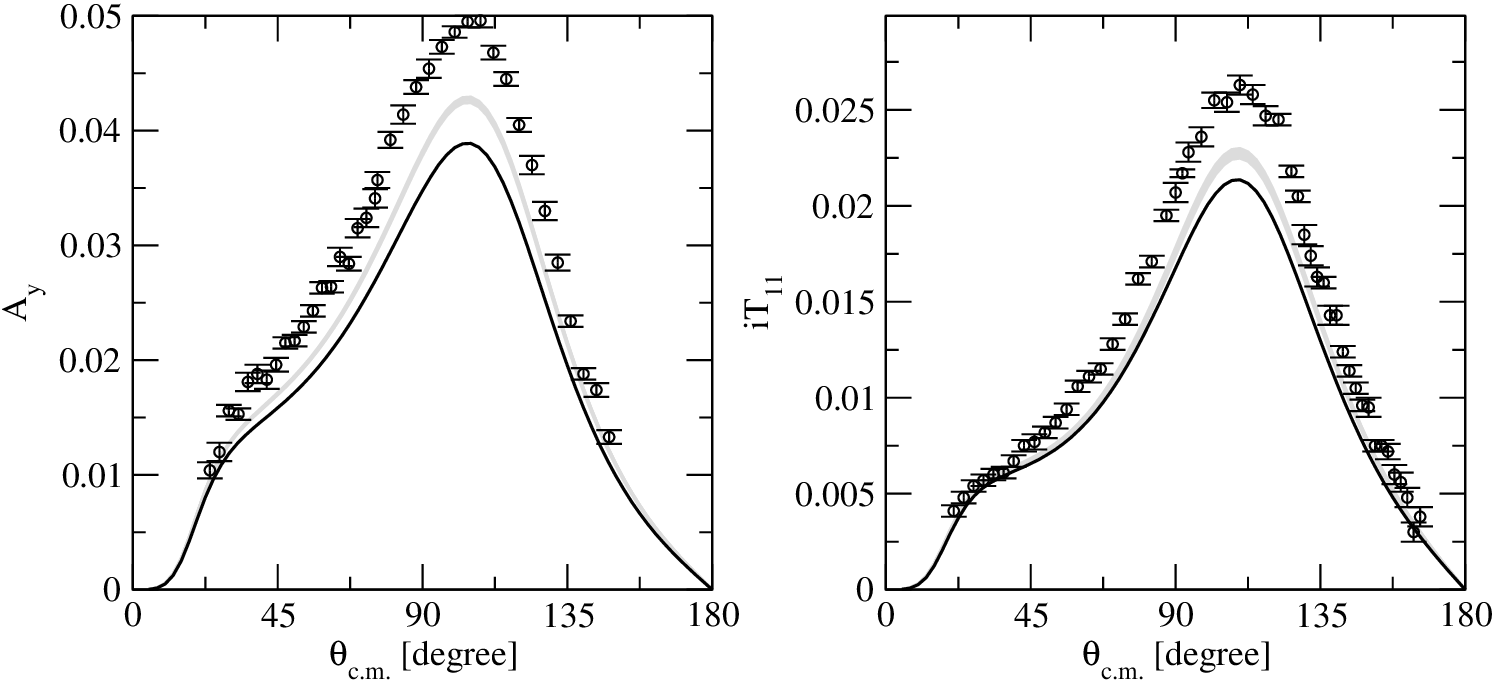}
\caption{The $p-d$ analyzing powers $A_y$ and $iT_{11}$ at $E_p=3$ MeV for the
AV18+N2LOL (grey band) and AV18+UR (solid line) models. Experimental data
are from Ref.~\protect\cite{shimizu}.}
\label{fig:fig2}
\end{figure}

In Ref.~\cite{kiev10b}
a comparative study of the aforementioned TNF models has been performed.
Let us briefly review their structure. From the general form
\begin{equation}
W= \sum_{i<j<k} W(i,j,k)  \;\; ,
\label{eq:wijk}
\end{equation}
a generic term can be decomposed as
\begin{equation}
W(1,2,3)= aW_a(1,2,3)+bW_b(1,2,3)+dW_d(1,2,3)+c_DW_D(1,2,3)+c_EW_E(1,2,3) \; .
\label{eq:w123}
\end{equation}
Each term corresponds to a different mechanism and has a different operatorial
structure.  The specific form of these terms in configuration space is:
\begin{equation}
\begin{aligned}
& W_a(1,2,3) = W_0(\bm\tau_1\cdot\bm\tau_2)(\bm\sigma_1\cdot \bm r_{31})
           (\bm\sigma_2\cdot \bm r_{23}) y(r_{31})y(r_{23}) \\
& W_b(1,2,3)= W_0 (\bm\tau_1\cdot\bm\tau_2) [(\bm\sigma_1\cdot\bm\sigma_2)
  y(r_{31})y(r_{23})  \\
 &\hspace{2cm} + (\bm\sigma_1\cdot \bm r_{31})
    (\bm\sigma_2\cdot \bm r_{23})(\bm r_{31}\cdot \bm r_{23})
  t(r_{31})t(r_{23}) \\
 & \hspace{2cm} + (\bm\sigma_1\cdot \bm r_{31})(\bm\sigma_2\cdot \bm r_{31})
  t(r_{31})y(r_{23}) \\
 &\hspace{2cm} + (\bm\sigma_1\cdot \bm r_{23})(\bm\sigma_2\cdot \bm r_{23})
  y(r_{31})t(r_{23})] \\
& W_d(1,2,3)=W_0(\bm\tau_3\cdot\bm\tau_1\times\bm\tau_2)
   [(\bm\sigma_3\cdot \bm\sigma_2\times\bm\sigma_1)y(r_{31})y(r_{23}) \\
  & \hspace{2cm}+ (\bm\sigma_1\cdot \bm r_{31})
    (\bm\sigma_2\cdot \bm r_{23})(\bm\sigma_3\cdot\bm r_{31}\times \bm r_{23})
  t(r_{31})t(r_{23}) \\
 & \hspace{2cm}
  + (\bm\sigma_1\cdot \bm r_{31})(\bm\sigma_2\cdot \bm r_{31}\times\bm\sigma_3)
  t(r_{31})y(r_{23}) \\
 &\hspace{2cm} + (\bm\sigma_2\cdot \bm r_{23})(\bm\sigma_3\cdot \bm r_{23}\times
  \bm\sigma_1) y(r_{31})t(r_{23})]\;\; ,
\end{aligned}
\end{equation}
with $W_0$ an overall strength.
The $b$- and $d$-terms are present in the three models whereas the $a$-term
is present in the TM' and N2LOL and not in URIX.
The last two terms in Eq.(\ref{eq:w123}) correspond to a two-nucleon (2N) contact term
with a pion emitted or absorbed ($D$-term) and to a three-nucleon (3N)
 contact interaction ($E$-term). Their local form, derived in Ref.~\cite{N2LO}, is
\begin{equation}
\begin{aligned}
& W_D(1,2,3)= W_0^D (\bm\tau_1\cdot\bm\tau_2) \{(\bm\sigma_1\cdot\bm\sigma_2)
  [y(r_{31})Z_0(r_{23})+Z_0(r_{31})y(r_{23})]  \\
 & \hspace{2cm} + (\bm\sigma_1\cdot \bm r_{31})(\bm\sigma_2\cdot \bm r_{31})
  t(r_{31})Z_0(r_{23}) \\
 &\hspace{2cm} + (\bm\sigma_1\cdot \bm r_{23})(\bm\sigma_2\cdot \bm r_{23})
  Z_0(r_{31})t(r_{23})\} \\
& W_E(1,2,3) = W_0^E(\bm\tau_1\cdot\bm\tau_2) Z_0(r_{31})Z_0(r_{23})  \,\, .
\end{aligned}
\end{equation}
The constants $W_0^D$ and $W_0^E$ fix the strength of these terms.
In the case of the URIX model the $D$-term is absent whereas
the $E$-term is present without the isospin operatorial
structure and it has been included as purely phenomenological, without
justifying its form from a particular exchange mechanism. The different
form for the profile functions of each model, $y(r)$, $t(r)$ and $Z_0(r)$
are given in Ref.~\cite{kiev10b}.
In that reference the strengths relative to the different terms have been varied in
order to reproduce, as close as possible, $B$($^3$H) and $B$($^4$He) and
$^2a_{nd}$. With these new parametrizations selected polarization observables
can been calculated and compared to the experimental data. In particular using
the N2LOL three-nucleon force a small improvement in the description of the
vector analyzing powers $A_y$ and $iT_{11}$ at low energies has been obtained. This
is shown in Fig.~\ref{fig:fig2} in which the predictions for $p-d$ $A_y$ and
$d-p$ $iT_{11}$ at $E_p=3$ MeV of the AV18+N2LOL model (grey band) is compared to
those ones of the AV18+UR model (solid line). From the figure we can observed that
the discrepancy has been appreciable reduced.

\begin{figure}
\includegraphics[width=14cm,height=8cm,angle=0]{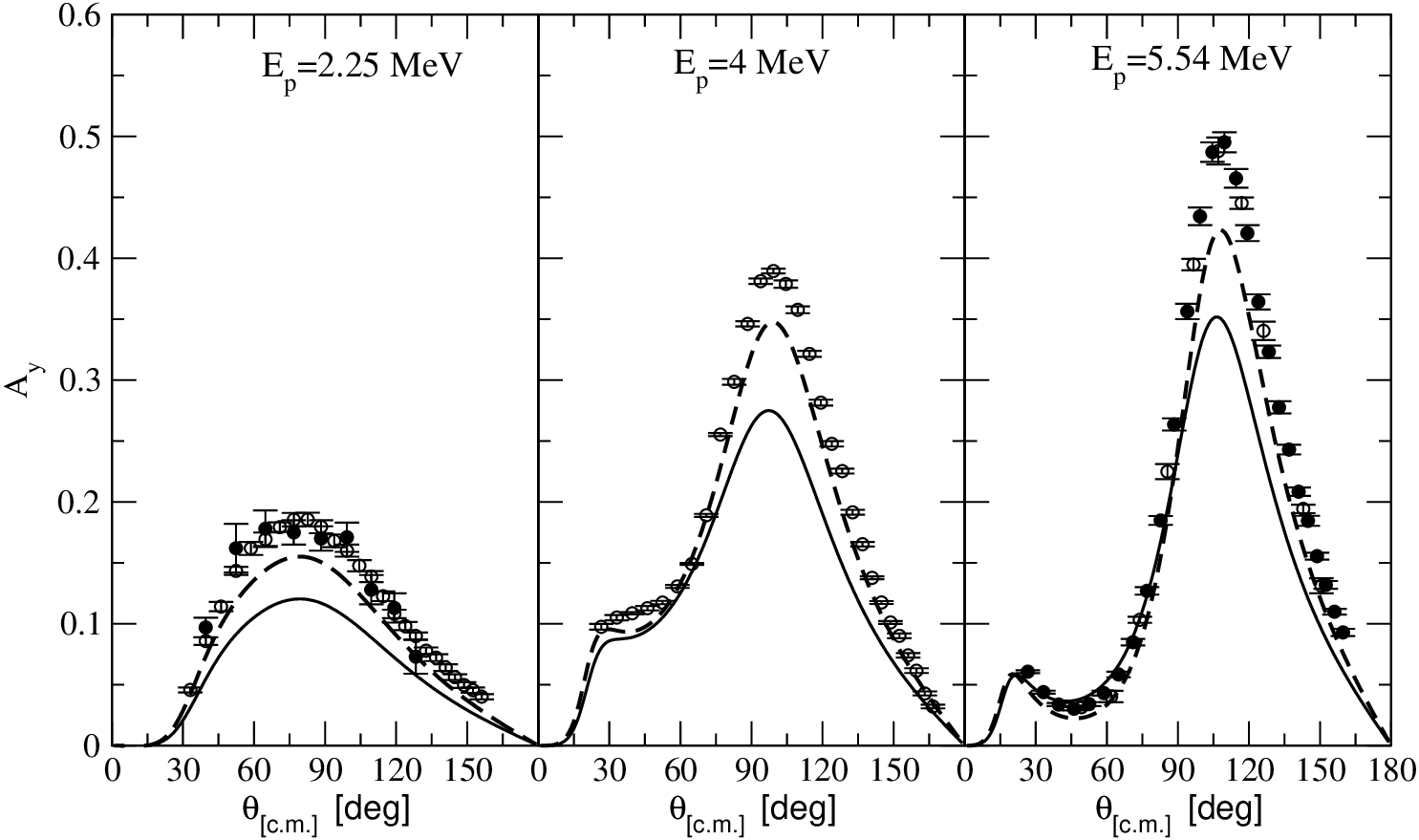}
\caption{The $p-^3$He analyzing powers $A_y$ at three energies for the
N3LO-Idaho+N2LOL (dashed line) and AV18+UR (solid line) models. Experimental data
are from Refs.~\protect\cite{alley,daniels}.}
\label{fig:fig3}
\end{figure}

The effects of the N2LOL TNF is much more evident in the four nucleon system.
In fact, in Fig.~\ref{fig:fig3} the $p-^3$He analyzing power $A_y$ is shown
at three energies using the N3LO-Idaho NN interaction plus the N2LOL TNF
(dashed line) and the AV18+UR model (solid line). We can see the big effect
produced by the inclusion of the N2LOL TNF model. It should be noticed that
the observable calculated using the N3LO-Idaho or the AV18
NN forces alone results close to the predictions of the AV18+UR model 
(see Ref.~\cite{viv09}), indicating that the improvement is given by the
inclusion of the N2LOL force. As in the three-nucleon system, this TNF
model considerable reduces the discrepancy obtained in the description
of this observable.

\section{Conclusions}
\label{sec:conclu}

Two different aspects of the description of few-nucleon systems have been
discussed. Firstly, scattering states below the deuteron breakup threshold
has been calculated using bound state like wave functions. The starting point
in this analysis was the integral relations recently derived from the KVP.
Finally an analysis of the effects of TNF models has been briefly discussed
in the vector analyzing powers in $p-d$ and $p-^3$He scattering. In particular
it was shown that the inclusion of the N2LOL TNF appreciable improves the
description of those observables. Further studies on these subjects are at
present in progress.

\begin{acknowledge}
The results presented in this work have been obtained in collaboration with
C. Romero-Redondo and E. Garrido (CSIC), P. Barletta (UCL) and my colleagues
in Pisa, M. Viviani, L. Girlanda and L.E. Marcucci. 
\end{acknowledge}

\end{document}